\documentclass[reprint,amsmath,amssymb,showkeys,aps,prb]{revtex4-1}
\usepackage{natbib}
\usepackage{graphicx}
\usepackage{dcolumn}
\usepackage{bm}
\usepackage{hyperref}

\begin{document}

\title{Superlubric-Pinned Transition in Sliding Incommensurate Colloidal Monolayers}

\author{
        Davide Mandelli$^1$, 
        Andrea Vanossi$^{2,1}$, 
        Michele Invernizzi$^{3}$, 
        S. V. Paronuzzi Ticco$^{1}$, 
        Nicola Manini$^{3,1,2}$, 
        Erio Tosatti$^{1,2,4}$
       }

\affiliation{
$^1$ International School for Advanced Studies (SISSA),
     Via Bonomea 265, 34136 Trieste, Italy \\
$^2$ CNR-IOM Democritos National Simulation Center,
     Via Bonomea 265, 34136 Trieste, Italy \\
$^3$ Dipartimento di Fisica, Universit\`a degli Studi di Milano, 
     Via Celoria 16, 20133 Milano, Italy \\
$^4$ International Centre for Theoretical Physics (ICTP),
     Strada Costiera 11, 34014 Trieste, Italy
}

\date{\today}
\begin{abstract}

Two-dimensional (2D) crystalline colloidal monolayers sliding over a
laser-induced optical lattice providing the periodic ``corrugation''
potential recently emerged as a new tool for the study of friction
between ideal crystal surfaces.
Here we focus in particular on static friction, the minimal sliding
force necessary to depin one lattice from the other.
If the colloid and the optical lattices are mutually commensurate, the
colloid sliding is always pinned by static friction; but when they are
incommensurate the presence or absence of pinning can be expected to
depend upon the system parameters, like in one-dimensional (1D) systems.
If a 2D analogy to the mathematically established Aubry transition of
one-dimensional systems were to hold, an increasing periodic corrugation
strength $U_0$ should turn an initially free-sliding, superlubric
colloid into a pinned state, where the static friction force goes from
zero to finite through a well-defined dynamical phase transition.
We address this problem by the simulated sliding of a realistic model 2D
colloidal lattice, confirming the existence of a clear and sharp
superlubric-pinned transition for increasing corrugation strength.
Unlike the 1D Aubry transition which is continuous, the 2D transition
exhibits a definite first-order character, with a jump of static
friction.
With no change of symmetry, the transition entails a structural
character, with a sudden increase of the colloid-colloid interaction energy,
accompanied by a compensating downward jump of the colloid-corrugation energy.
The transition value for the corrugation amplitude $U_0$ depends upon
the misalignment angle $\theta$ between the optical and the colloidal
lattices, superlubricity surviving until larger corrugations for angles
away from the energetically favored orientation, which
is itself generally slightly misaligned, as shown in recent work.
The observability of the superlubric-pinned colloid transition is
proposed and discussed.
\end{abstract}
\pacs{pacs}
\maketitle
\section{Introduction}
\label{sec:Intro}
\begin{figure}
 \begin{center}
 \includegraphics[angle=0, width=0.45\textwidth]{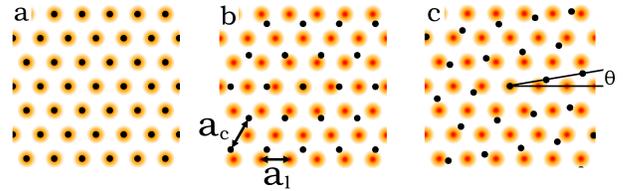}
 \caption{(Color online)
   (a) Schematics of a $\rho =1$ commensurate interface between a
   colloidal monolayer (black dots) and a triangular optical lattice
   potential (light spots correspond to the minima).
   (b) A colloidal monolayer of lattice spacing $a_{\rm c}$ with a
   mismatched triangular potential of periodicity $a_{\rm l}=\rho a_{\rm
     c}$.
   The two lattices are mismatched but aligned.
   (c) Colloidal monolayer with a small misfit angle $\theta$ with
   respect to the mismatched substrate.
   This kind of rotation generally lowers the energy and is a likely
   equilibrium feature.\cite{mandelli2015}
 }
 \label{fig_interfaces}
 \end{center}
\end{figure}

The great progress of nanofriction of the last decades has enabled
increasing insight into the microscopic nature of
friction.\cite{BOBOOK,vanossiRevModPhys,urbakhNatMat}
In particular, the sliding between contacting crystalline surfaces has
become of higher interest after the manipulation of nanoscale-sized
objects revealed how the atomistic features of an interface may play the
ultimate role in determining its tribological properties. 
Fresh interest is now generated by artificial but highly instructive systems represented
by 2D colloidal monolayers sliding in an optical lattice, the main system which we use 
as a working example in this study.~\cite{Bohlein12}
Two atomically flat and perfect crystal surfaces in contact are
commensurate when the mismatch ratio $\rho$ between their lattice
parameters is rational.
Perfect matching ($\rho=1$) between two lattices of same symmetry
represents the simplest situation, where the atoms of one crystal
lattice (the slider) perfectly fit the minima of the two-dimensional
``corrugation'' periodic potential landscape generated by the other
lattice (see Fig.~\ref{fig_interfaces}a).
In this case a finite and large external force is
required to dislodge the slider atoms from the potential minima, so as
to nucleate, at finite temperature,\cite{reguzzoni2010,Hasnain13} the
onset of sliding.

A different situation arises when the periodicities of the two
contacting lattices are mismatched, as is realized either by
a value $\rho\neq 1$, or else by a mutual rotation of two
lattices, as shown schematically in Fig.~\ref{fig_interfaces}b.
The present study concerns interfaces that are incommensurate.
A situation far more common in nature, it also naturally exhibits
more interesting tribological features than commensurate
ones.\cite{Vanossi12PNAS}
We focus especially on static friction, the minimal sliding force necessary
to depin one lattice from the other.
The main issue we address here is the nature of static friction against
sliding of a 2D lattice onto a 2D incommensurate corrugation.

The reference point for our understanding is the 1D case, epitomized by
the so-called Frenkel-Kontorova (FK) model, consisting of a harmonic
chain of classical point particles in a static sinusoidal potential, an
idealized system whose physics has been very widely studied and
understood.\cite{braun_kivshar}
In 1D, the incommensurability between interparticle spacing and sinusoidal
potential wavelength gives rise to a potential-induced deformation of the
chain particle position, that can be described by a deformation of the
chain's phase $\Phi(x)$ relative to that of the potential, from the
non-interacting straight line of slope $(\rho - 1)$ to a staircase with the
same mean slope but formed by nearly horizontal, approximately constant
phase terraces.
These commensurate domains are separated by steps in the relative phase,
often called solitons (or kinks) and antisolitons (or antikinks), where
most of the misfit stress is concentrated.

This 1D system is known to display a remarkable dynamical phase transition,
first described by Aubry.\cite{aubry1983}
When the corrugation is weak, below a critical magnitude, when the soliton
widths are comparable to or larger than that of terraces and the overall phase
modulation is small, the FK chain exhibits no pinning, the static friction is zero.
In this regime one finds particles at all potential values,
and all possible phases of the chain particles are accessible
in dynamics. Since the total potential energy is independent of the phase, $\delta
E/\delta \Phi$ = 0, the chain as a whole can be displaced by any
infinitesimal force. This absence of static friction has been dubbed 
``superlubricity''.\cite{hirano1993}
However, above a critical corrugation magnitude, whose value depends on
precise parameters and incommensurability, the chain develops a nonzero
static friction through a continuous ``Aubry'' phase transition.
The essence of the Aubry transition is that the probability to find a
particle exactly at a potential maximum drops mathematically to zero (of
course at $T=0$).
Thus, even if $\delta E/\delta \Phi$ is still zero, this new dynamical
constraint limits phase-space accessibility, breaking ergodicity in a
way similar to a glass transition, with the onset of static friction and
of chain pinning against free sliding.
This transition is of a dynamical nature, so that even if the step-terrace
deformation of the chain's phase $\Phi(x)$ becomes more marked in the
pinned state, it still remains similar in nature to the superlubric one.
Although suggestive, the static structure difference between the two
does not reveal the essence of the transition.
A more relevant signature is provided by the sliding friction.
In the superlubric state, sliding of the chain occurs under any applied
force however small, and can be envisaged as a state of flow of the solitons.
Soliton motion causes dissipation, increasing with speed
in correspondence with increasing emission of phonons.
In the pinned state, chain sliding occurs only once the static friction
force is overcome, at which point the sliding dynamics becomes generally 
quite different from the superlubric state.

In view of this well-known 1D case, it is physically clear that a qualitatively
similar state of affairs should also occur in 2D incommensurate sliders.
That is actually a case much more relevant in practice, where one expects free,
superlubric sliding for weak corrugation, and pinning with stick-slip
for strong corrugation.
Superlubricity is documented in several real 2D systems.
Very small values of the static friction force $F_s$ were observed for a
rotationally misaligned graphene flake sliding over
graphite,\cite{Dienwiebel04} the pinned configuration corresponding to
the aligned and therefore commensurate case.
Superlubricity and consequent ultra-low dynamic friction has also been
demonstrated or implied in a number of cases, such as also telescopic
sliding among carbon nanotubes,\cite{zhang2013,bocquet2014} in cluster
nanomanipulation studies,\cite{dietzel2008,dietzel2013} and in the
sliding of rare-gas adsorbates.\cite{Woll,pierno2015}
However, there is no experimental case so far exhibiting a clear
superlubric-pinned transition among 2D incommensurate sliders under
fixed geometrical conditions.
An experimental system of fresh relevance where that goal might be
pursued is represented precisely by two-dimensional crystalline colloidal
monolayers sliding over a laser-induced periodic corrugation potential.
Measurements\cite{Bohlein12} and computer simulations\cite{Vanossi12PNAS} of 
colloidal monolayers interacting with a laser-induced periodic potential demonstrated 
the possibility to attain negligible values of the static friction force in incommensurate
geometries.
Therefore a study of the transition from the superlubric to the pinned
state, done at constant lattice mismatch, and at constant mutual
alignment is potentially feasible for sliding colloid monolayers.
These systems are excellent candidates as they allow to change all the
relevant parameters of the interface: the corrugation strength $U_0$ and 
the colloidal crystal lattice parameter $a_{\rm c}$, which sets the mismatch 
ratio $\rho = a_{\rm l}/a_{\rm c}$ given the optical potential periodicity 
$a_{\rm l}$, fixed by the laser.
The mutual alignment or misalignment between the colloid and the optical
lattices is decided by energetics.
As was established by a recent study, a small misalignment generally
prevails, directly detectable by the colloid moir\'e pattern.\cite{mandelli2015}
This kind of misalignment, depicted in Fig.~\ref{fig_interfaces}c, is
well established in surface-adsorbed rare-gas monolayers.\cite{novaco1977, Shiba79, Shiba80}

The question we address here, only marginally touched
upon so far\cite{braun_kivshar} is whether at constant mismatch
the weak- and the strong-corrugation regimes will
again or will not be separated by a well-defined superlubric--pinned
phase transition of the same type as that described by Aubry in 1D --
the only case where a mathematical treatment has been worked out.
Colloid monolayers offers the ideal chance to verify this point if not
analytically at least numerically in a realistic and relevant case.

For a 2D triangular crystal monolayer interacting
with a mismatched rigid corrugation periodic potential the 1D
soliton-terrace staircase is replaced by a 2D superlattice of nearly
commensurate domains.
The domains are separated by soliton (antisoliton) misfit dislocation
lines, depending on the specific value of $\rho$, and of the misalignment
angle.
In triangular symmetry, three families of soliton lines, oriented at
$\pm$120 degrees from one another, will generally coexist.
The coincidence plot between the two lattices (monolayer and periodic
corrugation) realizes a moir\'e pattern, consisting of a patchwork of
nearly commensurate, roughly hexagonal domains separated by 
the soliton lines, which concentrate the mismatch
of the two lattices.\cite{Vigentini14}
Like in 1D, the sliding of the 2D monolayer enacts a ''flow'' of the
solitons in the moir\'e pattern.
A spontaneous angular misalignment, which when present lowers the energy
by increasing the interdigitation of the two lattices, strongly modifies
that pattern and correspondingly increases friction.\cite{mandelli2015}

The main questions which we plan to address here are:
(a) Is there a sharp superlubric-pinned transition in the sliding of
incommensurate colloid monolayers for increasing corrugation strength?
(b) Is this 2D transition continuous and critical (as in 1D), or first order?  
(c) Does the spontaneous misalignment affect the transition, and when so, to
what effect?

Here we present simulations for a model incommensurate monolayer which
show that: (a) Superlubricity is replaced by pinning through a sharp
transition. (b) That transition is, at least for the set of 
parameters used, of first order rather than continuous. (c) Spontaneous
rotations very definitely affect the transition point, decreasing its
critical value of corrugation $U_0$.

The paper is organized as follows.
In Sect.~\ref{sec:Mod&Prot} we introduce the model and the protocols
adopted in the molecular dynamics simulations.
In Sect.~\ref{sec:ResLLM} we present results which indicate a
first-order structural transition taking place in the colloid monolayer
as a function of increasing corrugation.
Section~\ref{sec:ResFs} discusses the coincidence between this
structural transition and the first-order onset of a static friction
force $F_s$.
In Sect.~\ref{sec:ResThetaDep} we describe the dependence of the
static properties of the monolayer on the misalignment angle.
Section~\ref{sec:Concl} contains a discussion of possible experimental
verifications, as well as our final remarks and conclusions.

\section{Model and simulations}
\label{sec:Mod&Prot}
We describe the colloidal dynamics using the same model and parameters
as in Refs.~\onlinecite{Vanossi12PNAS,mandelli2015}.
In short, the 2D layer of colloidal particles is represented by
screened-Coulomb repulsive, classical point-like particles, moving in 2D
in a periodic potential (the corrugation) of amplitude $U_0$,
periodicity $a_{\rm l}$, and triangular symmetry.
The overdamped dynamics of these particles is generated by integrating
$T=0$ damped equations of motion, with a large viscous damping
coefficient $\eta=28$.
All results are expressed in terms of the system of units defined in
Table~I of Ref.~\onlinecite{mandelli2015}.
Simulations were performed adopting 2D periodic boundary conditions (PBC). 
Any desired misalignment angle between the corrugation and the colloidal
lattice is implemented by means of a suitably chosen supercell, built
using a standard procedure as follows.\cite{trambly2010}
The two lattices are defined by the pairs of primitive vectors
$\mathbf{a}_1=a_{\rm l}(1,0)$, $\mathbf{a}_2=a_{\rm l}(0.5,\sqrt{3})$,
and 
$\mathbf{b}_1=a_{\rm c}(\cos\theta,\sin\theta)$, $\mathbf{b}_2=a_{\rm
  c}(\cos(\theta+\pi/3),\sin(\theta+\pi/3))$.
The overall supercell-periodic structure can be realized if four
integers exist which satisfy the matching condition
$n_1\mathbf{a}_1+n_2\mathbf{a}_2=m_1\mathbf{b}_1+m_2\mathbf{b}_2$.
The ensuing supercell is that of a triangular lattice of size
$L=|m_1\mathbf{b}_1+m_2\mathbf{b}_2|$,
containing a total number of particles $N_p=m_1^2+m_1m_2+m_2^2$.
In practice we fix $a_{\rm c}=1$ and vary $n_{1,2}$, $m_{1,2}$ in search
of structures with a mismatch $\rho=a_{\rm l}/a_c\approx3/(1+\sqrt{5})$
-- close to the experimental values of Ref.~\onlinecite{Bohlein12} --
and $\theta$ near the desired value, with the obvious additional
constraint that the number of particles $N_p$ should not be too large.
We consider in practice the aligned configuration $\theta=0$, plus
the following misaligned configurations: $\theta\simeq 5^\circ$,
$\theta\simeq 10^\circ$ and $\theta_{\rm opt}\simeq 2.54^\circ$.
The latter is close to the (Novaco-McTague) equilibrium misalignment
angle $\theta_{\rm NM}\simeq 2.58^\circ$ predicted by weak-coupling
elastic theory.\cite{novaco1977,mandelli2015}
Table~\ref{tab_supercells} collects the adopted supercell parameters.
\begin{center}
 \begin{table}
 \renewcommand{\arraystretch}{1.3}
 \begin{tabular}{ | c | c | c | c | c | }
 \hline
 ($n_1$,$n_2$) & ($m_1$,$m_2$) & $\rho$ & $\theta$ & $N_p$ \\
 \hline
  (96,96) &  (89,89) & 0.92708333 & $0$    &  7,921 \\
 \hline
 (105,59) & (107,71) & 0.92705245 & $2.54^\circ$ & 20,701 \\
 \hline
  (96,72) & (100,54) & 0.92705012 & $5.07^\circ$ & 18,316  \\
 \hline
  (78,74) & (59,103) & 0.92705113 & $9.78^\circ$ & 17,332 \\
 \hline
 \end{tabular}
 \caption{\label{tab_supercells}
   Parameters of the four supercells adopted for the simulations in PBC
   (see text for definitions).
   The supercells are determined following the procedure outlined in
   the text.
   For $\theta=0$, the adopted mismatch $\rho=a_{\rm l}/a_{\rm c}=89/96$
   corresponds to the sixth approximation in the continued fraction
   expansion of $3/(1+\sqrt{5})$.
 }
 \end{table}
\end{center}

As mentioned earlier, the mismatch in this case produces a moir\'e
pattern which could be roughly described as a patchwork or superlattice
of small domains where colloids and corrugation are mutually (nearly)
commensurate, separated by a network of anti-soliton lines whose
thickness decreases with increasing corrugation strength.
From the experimental point of view, the underdense regime chosen here
for exemplification is better suited than the overdense case $\rho>1$
where local compressions may lead to buckling of particles out of the
plane, and where the energetics (not symmetrical with respect to
$\rho>1$) is less convenient.
The initial step of our study is to identify the distorted colloid
lattice configuration of lowest energy in presence of the corrugation.
For that, we carry out $T=0$ damped-dynamics simulations testing
different protocols.
In the first we increase the corrugation amplitude $U_0$ in steps
$\Delta U=0.018$ -- eventually reduced near transition critical points
-- then reversing the sign of $\Delta U$ and decreasing $U_0$ back to
zero in order to check for hysteresis across structural transitions.
The maximum values of the corrugation considered is $U_{\rm max}=1$.
In the second protocol we switch on abruptly the corrugation at given
values of $U_0$ with the monolayer initially either undistorted or in a
configuration previously fully relaxed at $U_{\rm max}$.
For each $U_0$ we eventually take the lowest-energy configuration.
Eventually the best configurations were always obtained using the first
protocol described above.

The depinning of the monolayer is studied by applying to each colloid a
driving force $F_d$, generally along a high-symmetry direction of the
laser substrate potential.
For a given value of $U_0$, the single particle barrier for the onset of motion is 
$F_{s1}=8\pi U_0/9a_{l}$. The external force is then increased adiabatically in 
steps $\Delta F=0.002$, much lower than the lowest value of $F_{s1}$ considered.
The duration is fixed in such a way that a single free particle would
slide by a total distance of $10\,a_{\rm c}$.
A colloid configuration is classified as sliding, when the procedure
produces a final center-of-mass displacement $\Delta x_{\rm cm}\gtrsim
2\,a_{\rm c}$.
\begin{figure}
 \begin{center}
 \includegraphics[angle=0, width=0.45\textwidth]{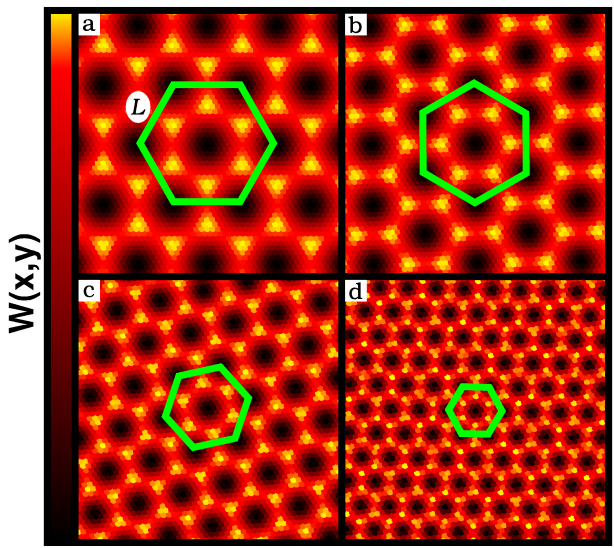}
 \caption{\label{fig_moire}(Color online)
   Examples of the moir\'e patterns obtained at $\rho\approx0.927$ for
   misfit angles (a) $\theta=0$, (b) $\theta_{\rm
     opt}\approx2.54^\circ$, (c) $\theta\approx5^\circ$, and (d)
   $\theta\approx10^\circ$.
   A small portion of the simulation supercells is displayed, containing
   an undistorted monolayer.
   The particles' colors reflect the underlying corrugation potential
   $W(x,y)$: dark for potential minima, bright for maxima.
   According to Eq.~\eqref{eq_psi}, at $\theta\simeq 2.54^\circ$ the moir\'e
   orientation is $\psi\simeq 30^\circ$.
   As $\theta$ increases the superstructure periodicity shrinks rapidly
   and rotates all the way to $\psi \simeq 60^\circ$ (c).
 }
 \end{center}
\end{figure}

Preliminary to presenting our results, it is useful to recall a few
geometrical concepts.
A given mismatch $\rho$ and misalignment angle $\theta$ induce a
specific moir\'e pattern of the colloidal crystal.
The moir\'e superstructure of an undistorted monolayer is entirely
described by its overall periodicity $L$ and its orientation angle
$\psi$ relative to the substrate.
Similar to the beats of two sinusoids in one dimension, the values of
$\psi$ and $L$ are governed by the difference $\mathbf{G}-\mathbf{G}'$
between the smallest reciprocal lattice vectors of the monolayer and
optical potential.
The relations connecting $\rho$ and $\theta$ to $L$ and $\psi$ are given
for example in Ref.~\onlinecite{bohr1992}.
Here we recall the angular relation
\begin{equation}
\label{eq_psi}
\cos{\theta}=\rho^{-1}\sin^2{\psi}+\cos{\psi}\sqrt{1-\rho^{-2} \sin^2{\psi}} \,,
\end{equation}
which, for $\rho \simeq 1$ produces a rapid reorientation of the moir\'e
pattern for small changes of $\theta \simeq 0$, as visualized in
Fig.~\ref{fig_moire}.

\section{Results: the transition is structural}
\label{sec:ResLLM}
\begin{figure}
 \begin{center}
 \includegraphics[angle=0, width=0.45\textwidth]{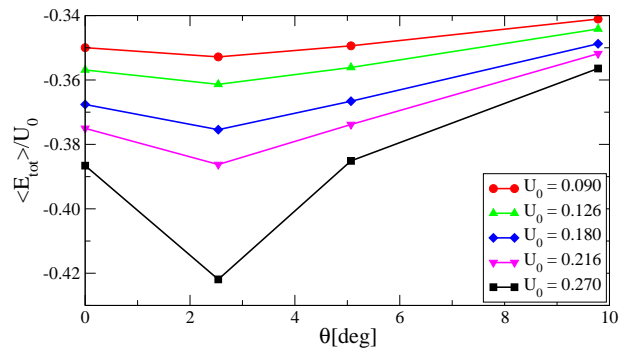}
 \caption{(Color online)
   The total energy $\langle E_{\rm tot} \rangle$ per particle of the
   colloidal monolayer optimized in PBC, for a few misalignment angles $\theta$.
   Curves correspond to increasing corrugation amplitude $U_0 \approx
   0.1-0.3$.
   Energies are measured relative to that of the perfect crystal at
   $U_0=0$, and are scaled with respect to $U_0$ for better comparison.
 }
 \label{fig_Etheta}
 \end{center}
\end{figure}

We begin by describing the evolution of the ground state
configuration of our model underdense colloidal monolayer as a function
of increasing corrugation strengths $U_0$, obtained at fixed
incommensurability $\rho\simeq 0.927$ and for a grid of angular
misalignment angles $\theta$.
Figure \ref{fig_Etheta} shows the total potential energy per particle
$\langle E_{\rm tot}\rangle$ as a function of $U_0$ and for several small $\theta$ values.
The lowest energy is seen to occur at a small but finite $\theta_{\rm opt}$.
This is a small but significant misalignment angle fairly close to the
theoretical, weak-coupling misalignment $\theta_{\rm NM}$ suggested by
the theory of orientational epitaxy.\cite{novaco1977,Shiba79,Shiba80}
The detailed  angular dependence of total colloid energy in the optical
lattice was investigated in Ref. \onlinecite{mandelli2015}.
The  mainly longitudinal deformation for  $\theta=0$,  is replaced in the optimal
misalignment $\theta_{\rm opt}$,  by energetically cheaper
largely transverse deformations.
This optimal misalignment angle is $\theta_{\rm opt} = 2.54^\circ$, to a
good approximation independent of corrugation $U_0$.
\begin{figure}
 \begin{center}
 \includegraphics[angle=0, width=0.45\textwidth]{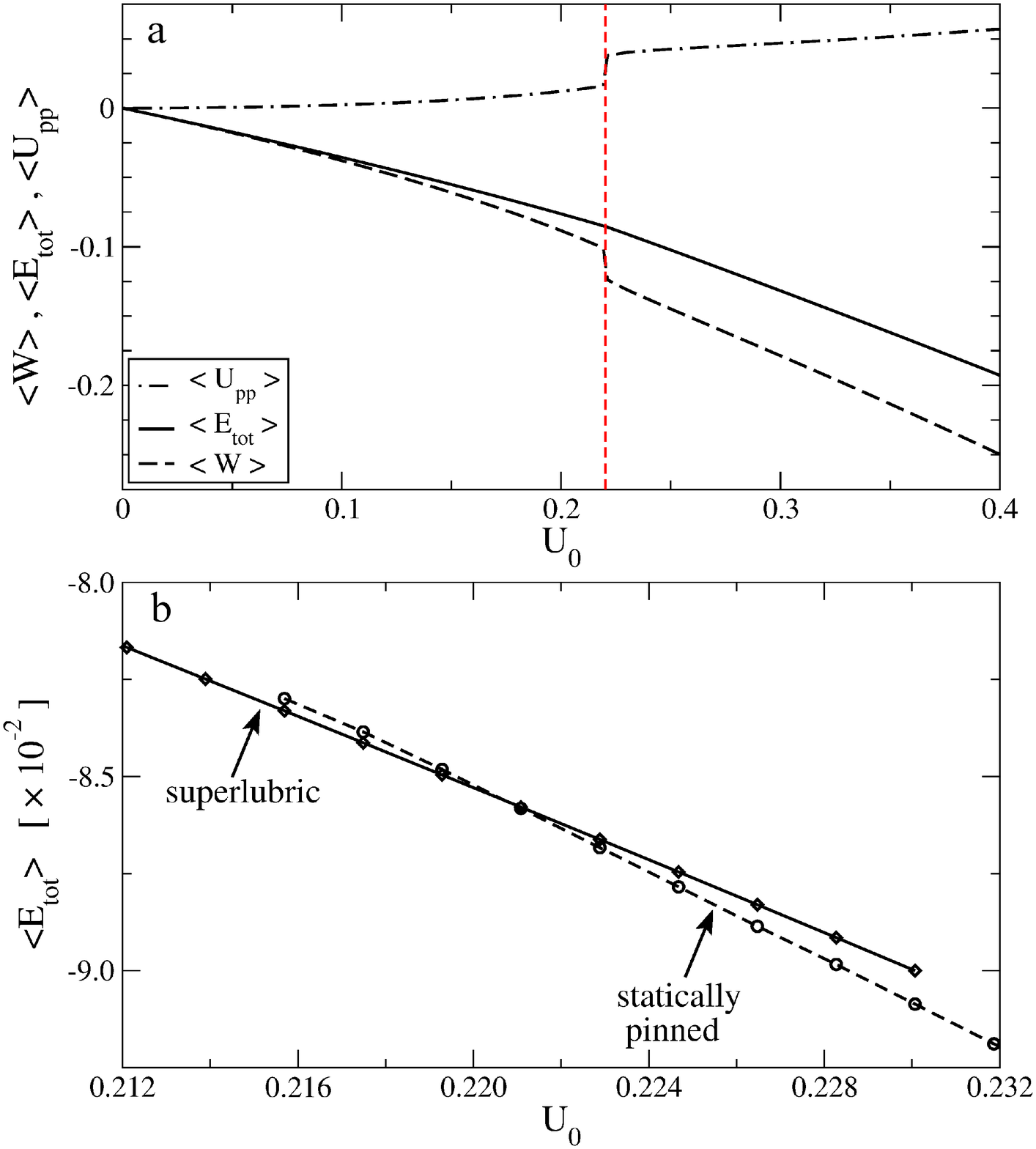}
 \caption{\label{fig_OptNRG}
 (Color Online)
   (a) The total potential energy $\langle E_{\rm tot}\rangle$ per particle, its
   periodic-potential (corrugation) contribution $\langle W\rangle$, and interparticle
   contribution $\langle U_{\rm pp}\rangle$ as a function of the corrugation $U_0$ for the
   optimally misaligned colloidal monolayer  ($\theta=\theta_{\rm opt} =
   2.54^\circ$).
   The sudden drop of $\langle W\rangle$ and the corresponding jump of $\langle U_{\rm pp}\rangle$
   signal a first-order transition, taking place at constant
   incommensurability and misalignment angle.
   As will be shown later, the weak-corrugation phase is unpinned and
   superlubric, the strong-corrugation phase is pinned by a finite
   static friction.
   (b) The intersection between the total energy branches
   representing the superlubric (small $U_0$) and statically pinned
   (large $U_0$) structures.
   Like in all first-order transitions, each phase survives in metastable
   state beyond the transition point.
 }
 \end{center}
\end{figure}

\begin{figure}
 \begin{center}
 \includegraphics[angle=0, width=0.45\textwidth]{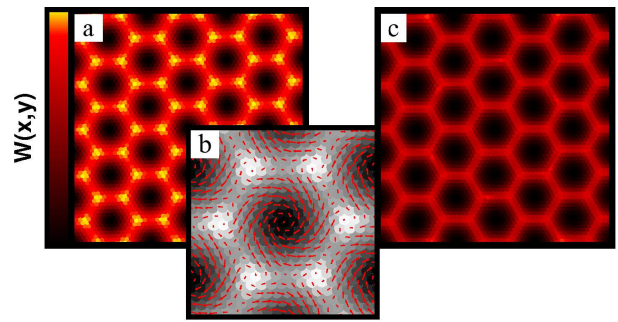}
 \caption{\label{fig_2DAubry}(Color online)
   The moir\'e patterns of the fully-relaxed incommensurate colloid
   monolayer at $\theta_{\rm opt}=2.54^\circ$.
   A small portion of the simulation supercell is shown.
   Each dot represents a particle, colored according to the local
   corrugation potential $W(x,y)$: dark for potential minima, bright for
   maxima.
   (a) The optimal weak-corrugation configuration ($U_0=0.108$).
   (b) Zoomed-in region with arrows highlighting the particles'
   displacements (magnified $20\times$) from the ideal triangular lattice
   to the fully-relaxed configuration.
   Vortex-like displacements tend to locally rotate the domains back
   into registry with the substrate.
   (c) The optimal strong-corrugation state ($U_0=0.395$).
   The hexagonal domains appear now fully commensurate, and separated by
   sharp domain walls.
   Here most colloids fall very close to a potential minimum.
 }
 \end{center}
\end{figure}

We wish to understand now what happens to the colloid structure, energy,
and eventually to the static friction as a function of $U_0$ at 
$\theta = \theta_{\rm opt}$.
For that purpose, we consider that the total potential energy per particle 
$\langle E_{\rm tot}\rangle = E_{\rm tot}/N_p$ is composed of two separate contributions:
\begin{equation}
\label{E_tot}
\langle E_{\rm tot}\rangle = \langle W\rangle + \langle U_{\rm pp}\rangle \,,
\end{equation}
where $\langle W\rangle$ is the average colloid-optical lattice interaction energy, and
$\langle U_{pp}\rangle$ the average colloid-colloid repulsion.
Figure~\ref{fig_OptNRG}a displays the two contributions to the total
energy $\langle E_{\rm tot}\rangle$.
The total energy evolution is smooth until $U_0=0.22$ where we observe a
sudden decrease in $\langle W\rangle$ accompanied by an increase of the
repulsive term.
This indicates a first-order structural transition in the monolayer.
Figure~\ref{fig_2DAubry}a displays the $\theta_{\rm opt}$ moir\'e
patterns of the relaxed monolayer for $U_0=0.108$, a weak corrugation
well below the first-order transition.
For this weak corrugation the triangular colloidal crystal is only
slightly affected by the substrate potential.
Panel \ref{fig_2DAubry}b illustrates the corresponding colloid
displacements relative to the rigid unperturbed crystal ($U_0=0$).
The misalignment of the colloidal crystal over the lattice corrugation
is only slightly modulated by a vortex-like distortion reflecting the
shear nature of the solitons, whose width is larger than their
separation.
Panel \ref{fig_2DAubry}c portrays the moir\'e pattern for $U_0=0.395$, a
strong corrugation well above the transition.
A hexagonal superlattice of domains of colloids locally commensurate
with the corrugation are separated by ``atomically'' sharp antisolitons
lines.

\begin{figure}
 \begin{center}
 \includegraphics[angle=0, width=0.45\textwidth]{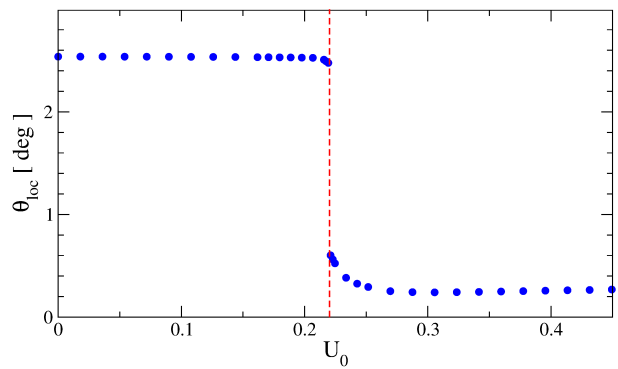}
 \caption{\label{fig_thetaloc} (Color online)
   The average local orientation of the monolayer $\theta_{\rm loc}$ (see
   text for definition) plotted as a function of corrugation.
   The vertical dashed line separates the weak-corrugation and 
   the strong-corrugation phases.
 }
 \end{center}
\end{figure}

It is also instructive to look at the average local orientation of the
monolayer (see Fig.~\ref{fig_thetaloc}), defined as
\begin{equation}\label{eq_thetaLOC}
\theta_{\rm loc}=\frac 1M \sum_{\langle i,j \rangle}
\bmod\!\left(\theta_{ij},\frac \pi3\right)
,
\end{equation}
where the sum is over all $M$ pairs $\langle i,j \rangle$ of nearest
neighbor particles, and $\theta_{ij}$ is the angle between the relative
position vector ${\bf r}_i-{\bf r}_j$ and the $x$-axis.
At the transition, the commensurate domains realign significantly with
the triangular potential.
Figure~\ref{fig_OptNRG}b shows a crossing of the total energies of the
weak-corrugation and the strong-corrugation phase at the transition
point, as a function of $U_0$.
A separate stability analysis shows that each phase survives as a
metastable state over a finite corrugation interval on the ``wrong''
side of $U_c$, eventually collapsing at its respective spinodal point.

\section{Results: Superlubricity and Static Friction}
\label{sec:ResFs}
\begin{figure}
 \begin{center}
 \includegraphics[angle=0, width=0.45\textwidth]{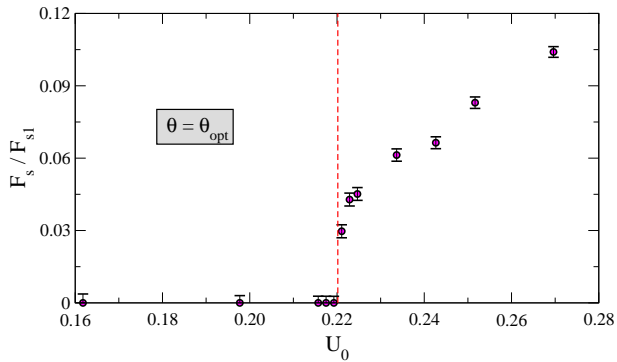}
 \caption{(Color online)
   The static friction force $F_s$
   of the $\rho\simeq 0.927$ colloidal monolayer, 
   normalized to the single-particle barrier $F_{\rm s1}$.
   The dashed line indicates the transition point corrugation $U_{\rm c}=0.22$ 
   where the colloid monolayer suddenly locks onto the corrugation.
   The error bar is defined by the force increment $\Delta F=0.002$
   used in the simulation protocol.
}
\label{fig_Fs}
\end{center}
\end{figure}

We now come to the static friction force $F_{\rm s}$ of the colloid
monolayer, the main interest of the present investigation.
It is obvious that the first-order structural transition uncovered in
the previous section must affect it.
Figure~\ref{fig_Fs} shows the static friction force $F_{\rm s}$ obtained
by simulation with $\rho\simeq 0.927$ and $\theta_{\rm opt}=2.54^\circ$
as a function of the corrugation strength $U_0$.
For weak corrugation $U_0<U_c$ the static friction is small below the error.
In this regime the colloid monolayer slides freely already at the
smallest applied force, demonstrating that the incommensurate state is
indeed superlubric.\cite{FOOT}
At the transition corrugation the static friction suddenly jumps to
$0.03 F_{\rm s1}$.
The monolayer becomes pinned, and the superlubricity is lost through a
first-order transition.
Unlike 1D, where the pinning transition was of second order, here it
coincides with the first-order structural transition.
\begin{figure}
 \begin{center}
 \includegraphics[angle=0, width=0.45\textwidth]{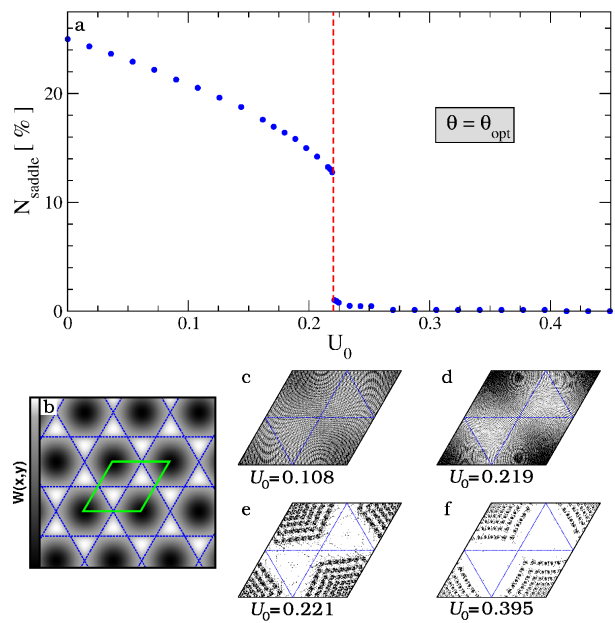}
 \caption{(Color online)
   (a) The fraction of colloidal particles which sample repulsive
   regions of the substrate potential, i.e.\ those positions where
   $W(x,y)$ exceeds the saddle energy, plotted as a function of the
   corrugation strength $U_0$.
   (b) The triangular potential $W(x,y)$. Dark and bright regions
   correspond respectively to minima and maxima, isolines are reported
   at the saddle-point value and a primitive cell is highlighted at the
   center of the density plot.
   (c-f) The positions of the particles reported inside one primitive cell
   of the substrate potential.
   Configurations are shown for different corrugation values: 
   (c) $U_0=0.108$ -- superlubric phase;
   (d) $U_0=0.219$ -- just below the transition;
   (e) $U_0=0.221$ -- just above the transition;
   (f) $U_0=0.395$ -- well above the transition.
 }
\label{fig_wrap}
\end{center}
\end{figure}

Leaving the transition order aside, it is easy to show that the
underlying reason for pinning in the present 2D case is still quite
similar to that in the 1D FK model, namely a collapse
of the dynamically-available phase space.
In 1D, that collapse was signaled by the simultaneous collapse of the
probability for any particle to occupy a local potential maximum.
In the present 2D case we can similarly investigate the
probability for colloids to occupy, in their optimal configuration, the
$(x,y)$ triangular region above the saddle-point 
among adiacent local minima of the corrugation potential.
If the analogy with 1D holds, a superlubric, ergodic state should
populate abundantly this region, while
in a pinned, broken-ergodicity state, the population probabilty at the
center of these regions should collapse to zero, and even remain zero
across a whole neighborhood of the maximum -- a finite ''disorder
parameter''.\cite{braun_kivshar}
Indeed, as shown in Fig.~\ref{fig_wrap}a, at the first-order
superlubric-pinned transition, these repulsive regions around the maxima
depopulate dramatically.
This is represented pictorially in Fig.~\ref{fig_wrap}c-f, reporting the
particles' positions folded in one primitive cell of the corrugation
lattice (see Fig.~\ref{fig_wrap}b).
These properties represent a 2D version of the disorder parameter
$\Psi$, which measures the width of the largest gap appearing in the
Hull function at the Aubry transition in the incommensurate 1D FK
model.\cite{braun_kivshar}
It should be noted that in the pinned colloid state the static friction force
remains well below the single-colloid limit value $F_{\rm s1}$, as shown
in Fig.~\ref{fig_Fs}.
The reason for this lower static friction is that of course the
monolayer sliding is still strongly collective, whereby the particles
surrounding a given particle attempting to cross a forbidden region
actually help by pushing it across the barrier.

\section{Results: Static Friction at Different Misalignments}
\label{sec:ResThetaDep}
\begin{figure}
 \begin{center}
 \includegraphics[angle=0, width=0.45\textwidth]{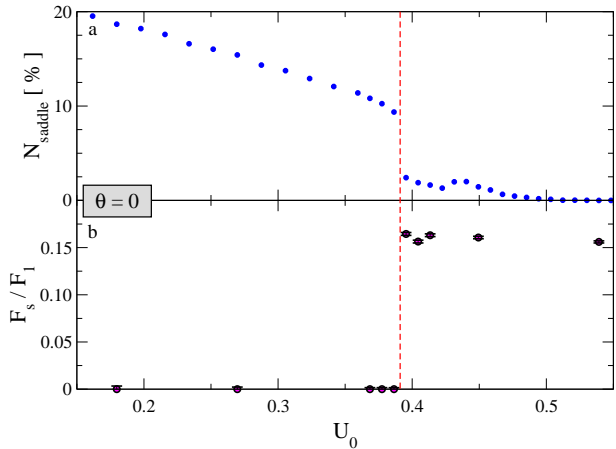}
 \caption{\label{fig_t0Fs}(Color online)
   Unrotated $\theta=0$ mismatched colloid monolayer,
   $\rho\approx0.927$.
   (a) Fraction of colloidal particles which populate repulsive regions
   of the corrugation potential (where $W(x,y)$ exceeds the saddle-point
   energy) plotted as a function of the corrugation strength $U_0$
   across the superlubric-pinned transition.
   The dashed line marks the transition point $\tilde U_c=0.391$.
   (b) Static friction force $F_s$, normalized to the depinning force
   $F_{\rm s1}$ of a single particle in the same periodic potential.
   Above the transition the monolayer locks onto the corrugation.
   The error bar is defined by the force increment $\Delta F=0.001$ used
   in the simulation protocol.
 }
 \end{center}
\end{figure}

An interesting aspect of the transition to the 2D locked state is its
dependence on the misfit angle $\theta$.
As discussed in Sect.~\ref{sec:ResLLM}, at the optimal orientation
$\theta_{\rm opt}$ the colloidal crystal deforms mainly via the softer
transverse modes, facilitating the interdigitation with the substrate.
If the monolayer is rotated away from $\theta_{\rm opt}$ its grip of the
corrugation weakens: as a result the superlubric state survives up to a
larger corrugation strength $U_0$.
As the simplest example, consider the aligned monolayer configurations
obtained at $\theta=0$.
\begin{figure}
 \begin{center}
 \includegraphics[angle=0, width=0.4\textwidth]{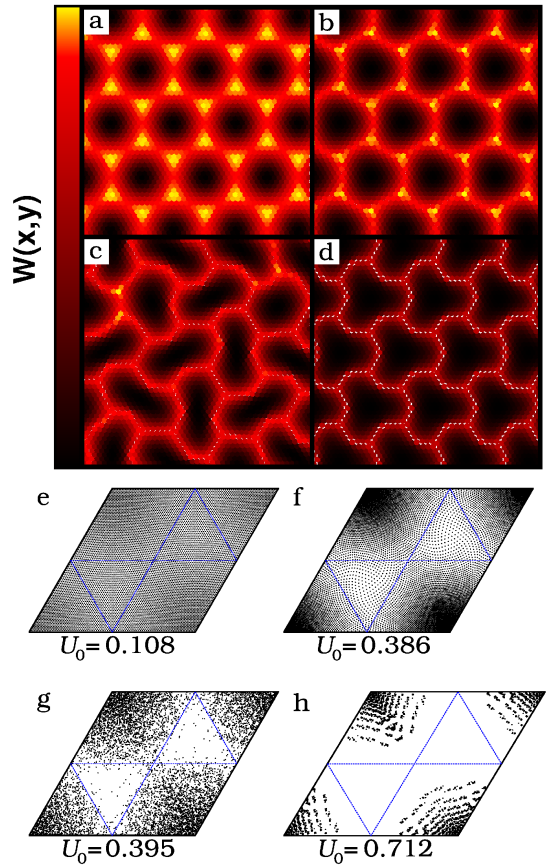}
 \caption{\label{fig_t0wrap} (Color online)
 Panels (a-d): the moir\'e patterns of the fully relaxed monolayer at
 $\theta=0$. 
 A small portion of the simulation supercell is shown.
 Each dot represents a particle, colored according to the local
 corrugation potential $W(x,y)$: dark for potential minima, bright for
 maxima.
 Panels (e-h): the particles' positions are reported into one primitive
 cell of the substrate triangular potential.
 Configurations are shown for the following values of the corrugation:
 (a,e) $U_0=0.108$ -- superlubric phase.
 (b,f) $U_0=0.386$ -- just below the transition.
 (c,g) $U_0=0.395$ -- just above the transition, where the energetically
 favored configuration consists of an arrangement of irregular domains.
 (d-h) $U_0=0.712$ -- a strong-corrugation phase corresponding to an
 arrangement of regularly-shaped registered domains.
 }
 \end{center}
\end{figure}

Figure~\ref{fig_t0Fs}a shows the evolution of the number of particles
sampling repulsive regions of the corrugation potential.
In this case too we identify a corrugation $\tilde U_c=0.391$ above which the monolayer
pins, again with a first-order transition.
As expected, the transition corrugation is larger -- almost a factor of
two -- than that at the optimal misalignment, $U_c=0.22$.
Figure~\ref{fig_t0Fs}b shows the static friction force $F_s$ across the
structural transition.
The aligned monolayer is superlubric up to $\tilde U_c$, where we
observe a sudden surge of the static friction $F_s$.
Actually in this metastable aligned case
the subsequent evolution of static friction becomes somewhat erratic.
This behaviour can be explained as follows.
A proliferation of energetically close metastable configurations -- each
characterized by its own static friction -- is expected\cite{aubry1983}
above the transition point.
While at $\theta_{\rm opt}$ all protocols produced configurations with
regularly-spaced hexagonal domains, at $\theta=0$ we observe several
both irregular and regular arrangements of registered domains, which
then happen to be energetically favored at different corrugation values
$U_0$ (see Fig.~\ref{fig_t0wrap}c,d).
Some fluctuations in $F_s$ are therefore expected when one also
considers that in incommensurate systems with many degrees of freedom,
states with lower energy do not necessarily display a higher static
friction force.\cite{BraunFuse}
This complexity disappears below $U_c$ where there always is a unique
structure, shown in Fig.~\ref{fig_t0wrap}a,b, respectively away and
close to the transition.
Summing up this part, the evolution of the monolayer structure as a
function of $U_0$ is highly sensitive to the misalignment angle.

\section{Discussion and Conclusions}
\label{sec:Concl}
Colloids in optical lattices are systems recently introduced,
potentially representing powerful tools which may contribute to
our understanding of friction between crystalline
surfaces.\cite{Bohlein12,Vanossi12PNAS,Hasnain13,Hasnain14}
Colloidal monolayers in an incommensurate lattice corrugation potential
are shown to represent ideal systems to study the superlubric-pinned
transition, expected as a function of increasing corrugation strength,
but never realized so far in any 2D system, under constant geometrical conditions.
In 1D it is well known that at this transition the onset of a disorder
parameter -- measuring a region of phase space inaccessible to particle
positions -- impedes the dynamics of free-sliding motion, giving rise to
static friction through a continuous phase
transition.\cite{aubry1983,braun_kivshar}
In the 2D case of colloids we predict a transition of similar character,
the main difference being now its first-order character.
The relative alignment and misalignment of the two lattices plays an
additional role.
We found vanishingly small static friction up to a critical value $U_c$
corresponding to a joint structural and dynamical transition of the
monolayer.
At the transition the colloidal crystal adapts to the periodic potential
by a local rotation (back to registry) of the nearly commensurate
domains forming the moir\'e of the incommensurate phase.

Similarly to the 1D case, pinning of the monolayer is due to a collapse
to zero of the number of particles sampling the most repulsive regions of
the periodic potential.
These poorly placed particles are those which are easily set into motion
in the superlubric phase.
Their disappearance is therefore responsible for the arisal of static
friction in the pinned state.
Experimentally, vanishingly small values of $F_s$ have been observed in
2D incommensurate structures,\cite{Vanossi12PNAS,Dienwiebel04} but the
realization and the analysis of the transition between superlubric and
pinned states is missing to our knowledge.
The present work suggests the possibility to pursue and investigate in
detail the superlubric-pinned transition by gradually increasing the
optical-lattice strength.
Additional phenomena may be accessed by forcing rotations and modifying
the natural misalignment of the colloidal crystal with respect to the
optical lattice.
Since the superlubric phase at $\theta\ne\theta_{\rm opt}$ survives up
to larger corrugation values, rotations of mismatched sliders initially
away from the optimal orientation might in fact end up in reentrant
pinning as soon as $\theta_{\rm opt}$ is hit, in a way similar to the
matched case ($\rho=1$, $\theta_{\rm opt}=0$) of a rotated graphene
flake sliding over graphite.\cite{filippov2008}

\acknowledgments
This work was mainly supported under the ERC Advanced Grant
No.\ 320796-MODPHYSFRICT, and partly by the Swiss 
National Science Foundation through a SINERGIA contract CRSII2\_136287,
by PRIN/COFIN Contract 2010LLKJBX 004, 
and by COST Action MP1303.
%

%%%%%%%%%%%BIBLIOGRAPHY%%%%%%%%%%%%%%%

%%%%%%%%%%%%%END DOCUMENT%%%%%%%%%%%%%%%

\begin{thebibliography}{99}
\bibitem{FOOT}
  Strictly speaking, since we are adopting PBC, our systems
  are commensurate and therefore ultimately pinned. A fine search at
  $U_0=0.216$, below and close to $U_c$, shows that the static friction
  force is indeed finite there, being $0.0005<F_s<0.001$. However, this
  ``pinning transition'' is marked by a sudden increase in $F_s$ by
  two-three orders of magnitude when increasing $U_0$ above the critical value $U_c$. 
  On the opposite side, for $U_0\ll U_c$, $F_s$ is so small that
  it is quite hard to measure.
\end{thebibliography}
\end{document}